# Universality of the self gravitational potential energy of any fundamental particle

Antonio Alfonso-Faus
Escuela de Ingeniería Aeroespacial
E.U.I.T. Aeronáutica, Universidad Politécnica de Madrid
Plaza Cardenal Cisneros 3, 28040 Madrid, Spain
e-mail: aalfonsofaus@yahoo.es

**Abstract.** - Using the relation proposed by Weinberg in 1972, combining quantum and cosmological parameters, we prove that the self gravitational potential energy of any fundamental particle is a quantum, with physical properties independent of the mass of the particle. It is a universal quantum of gravitational energy, and its physical properties depend only on the cosmological scale factor R and the physical constants $\hbar$ and c. We propose a modification of the Weinberg´s relation, keeping the same numerical value, but substituting the cosmological parameter H/c by 1/R.

**Keywords:** Cosmology, quantum mechanics, gravitational energy, cosmological scale factor, Weinberg´s relation.

In 1972 Weinberg [1] advanced a clue to suggest that large numbers are determined by both, microphysics and the influence of the whole universe. He constructed a mass using the physical constants G, $\hbar$, c and the Hubble parameter H. This mass was not too different from the mass of a typical elementary particle and is given by

$$m \approx (\hbar^2 H/Gc)^{1/3} \qquad (1)$$

We consider also a general elementary particle of mass m. The self gravitational potential energy $E_g$ of this quantum of mass m (and size its Compton wavelength $\hbar/mc$) is given by

$$E_g = Gm^2/(\hbar/mc) = Gm^3c/\hbar \qquad (2)$$

Combining (1) and (2) we can eliminate the mass m to obtain

$$E_g \approx \hbar H \qquad (3)$$

This expression has an important quantum-cosmological interpretation. We know today that the cosmological scale factor R is of the order of ct, t the age of the universe [2]. In this reference [2] the cosmological scale factor R is obtained in terms of the cosmological time t as

$$R(x)/R(1) = [2x/(3-x)]^{2/3} \qquad (4)$$

where $x = t/t_0$ is the dimensionless parameter for cosmological time in terms of the present age of the universe $t_0$. For $t = t_0$ we have $x = 1$. R(1) in (4) is the present value of the cosmological scale factor. The following fig. 1 gives the graphical plot of this cosmological scale factor R(x):

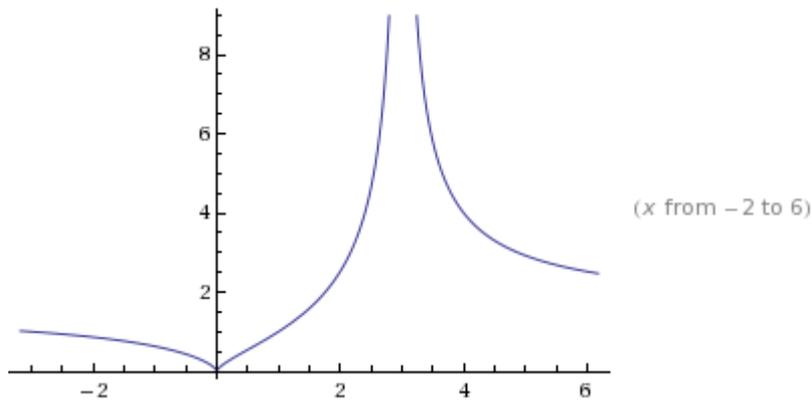

(x from −2 to 6)

Fig.1 Cosmological scale factor R(x)/R(1) versus time $x = t/t_0$

In this plot of R(x) versus x we see that there is an almost linear expansion law from x = 0 to about x = 1.4. A series Taylor expansion of (4) around x = 1 (today) gives

$$R(x)/R(1) = 1 + R'(1)/R(1)\,(x-1)/1! + R''(1)/R(1)\,(x-1)^2/2! + \ldots \quad (5)$$

Substituting the derivatives of R(x) from (4) into (5) we get

$$R(x)/R(1) \approx x + (x-1)^2/4 + O(x-1)^3 \quad (6)$$

The speed of expansion is then

$$R'(x)/R(1) \approx 1 + (x-1)/2 + O(x-1)^2 = (1+x)/2 + O(x-1)^2 \quad (7)$$

The acceleration is then

$$R''(x)/R(1) \approx 1/2 + O(x-1) \quad (8)$$

From the definition of the Hubble parameter H we have

$$H(x) = R'(x)/R(x) \approx [(1+x)/2]/x \quad (9)$$

Taking into account $dx = dt/t_0$ we have H(t) with its dimension 1/time

$$H(t) \approx 2/(t + t_0) \quad (10)$$

and from here we get for the present time $H(t_0) \approx 1/t_0$. Present values for the age of the universe give $t_0 \approx 1.37 \cdot 10^{10}$ years and therefore $H(t_0) \approx 71.4$ Km/(sec Mpc). This is well in the range of observations [3]. The Hubble radius for today is defined as $c/H(t_0) \approx c\,t_0$ and taking this radius as the present reference for the cosmological scale factor, i.e., c $t_0$ ≈ R(1) we get from (6), (7) and (8) to first order

$$R(t) \approx ct \; ; \; R'(t) \approx c \; ; \; R''(t) \approx 0 + c/(2 t_0) \qquad (11)$$

The expression for R´´(t) proves that there is a second order effect in the expansion of the universe, the acceleration c/2t. It has gone unnoticed up to a few years ago, [4] and [5]. The reason is here evident: it is a second order effect.

The central point of the result in (11) is that the ratio H/c is of the order of the inverse of R(t), at least close to the present time. In other words, following the same large number approach of Weinberg [1], we can use 1/R instead of H/c in its formulation. Substituting in (1) we get our revised Weinberg relation as

$$m \approx (\hbar^2/GR)^{1/3} \qquad (12)$$

Hence, substituting H/c ≈ 1/R in (3) we get the important result

$$E_g \approx \hbar c/R \qquad (13)$$

This is the Planck–Einstein equation. In 1923, Louis de Broglie postulated that the Planck constant $\hbar$ represents the product of the momentum and the quantum wavelength of not just the photon, but any particle. This was confirmed by experiments soon afterwards. Then we can interpret R in (13) as the wavelength of the quantum of gravitational energy. It is unlocalized, as the energy of the gravitational field [6]. And its momentum $p_g$ is then

$$p_g = E_g/c = \hbar/R \tag{14}$$

The equivalent expressions (3) and (13) for $E_g$ have very important implications. First, the elimination of the mass m of the elementary particle, any fundamental particle in fact, clearly makes the gravitational potential energy of any particle a universal expression. The presence of $\hbar$ in the expressions (3) and (13), and the cosmological parameters H or R, implies that, with the same aim as Weinberg had [1], we have obtained a universal relation between the physical constants, G, $\hbar$, and a cosmological parameter R (12). Second, the expression (13) can be interpreted as being a quantum of energy $E_g$, and it certainly has a gravitational nature as given by the initial concept used in (2): a gravitational potential energy. With these results we advance the following conjecture:

*The quantum of gravitational energy is given by the expression $E_g = \hbar c/R$, it has a wavelength of the order of the size of the universe, R, and a momentum $p_g = \hbar/R$. Its equivalent mass has a value $m_g = \hbar/cR \approx 10^{-66}$ grams, a figure found in many different instances in the scientific literature.*

The present knowledge of the composition of the universe is roughly 4% of normal baryon matter (stars and galaxies), 24% of non-baryon dark matter (inferred from its gravitational action on stars in galaxies, clusters of galaxies etc.) and 72% dark energy, from indirect arguments. All this mass (energy) must imply gravitation in some way, and therefore it represents an amount of gravitational energy corresponding to a total estimated mass for the universe of about M ≈ $10^{56}$ grams. Then, regardless of its origin, the total amount of gravity quanta should be MR/$\hbar$c ≈ $10^{122}$.

The gravitational potential energy of the universe is of the order of Mc² ≈ $10^{77}$ ergs. On the other hand the energy of the quantum of gravitation found here is $E_g$ = $\hbar$c/R ≈ $10^{-45}$ ergs. Dividing we get a number n ≈ $10^{122}$. This is a well known number. It may be linked to the entropy of the universe (with Boltzmann constant k = 1). It is also the order of magnitude of the discrepancy between the values of the cosmological constant Λ, as derived from cosmological information, and from the standard particle theory [7]. Here it may be thought as been a quantum number representative of the state of the universe as it is today. And there are n ≈ $10^{122}$ possibly entangled gravity quanta. This gives a pressure, and a corresponding energy density, that is well within the current known values for the seeable universe (with Hubble radius).

## Acknowledgement

I am thanking the owners of the Wolfram Mathematica Online Integrator that I have used to obtain the Figure 1 in this work.

REFERENCES

[1]. Weinberg, S., *Gravitation and Cosmology: Principles and Applications of the General Theory of Relativity*, page 619, John Wiley & Sons, Inc., (1972).

[2]. Alfonso-Faus, A., (2011) "Evidence for a disaggregation of the universe", arXiv: 1104.3781.

[3]. Hinshaw, G. et al. (WMAP Collaboration). (feb 2009). "Five-Year Wilkinson Microwave Anisotropy Probe Observations: Data Processing, Sky Maps, and Basic Results" (Table 7). *The Astrophysical Journal Supplement* **180** (2): 225–245. arXiv: 0803.0732.

[4]. Riess, A. G., et al. [Supernova Search Team Collaboration] *Astron. J.* **116**, 1009 (1998) [arXiv: astr-ph/9805201].

[5]. Perlmutter, S., et al. [Supernova Cosmology Project Collaboration], *Astrophys. J.* **517,** 565 (19999 [arXiv: astro-ph/9812133].

[6]. Misner, C.W., Thorne, K.S., and Wheeler, J.A., *GRAVITATION,* page 466: "WHY THE ENERGY OF THE GRAVITATIONAL FIELD CANNOT BE LOCALIZED", Freeman and Company, San Francisco, (1973).

[7]. Alfonso-Faus, A., "Artificial contradiction between cosmology and particle physics: the lambda problem", arXiv: 0811.3933 and *Astrophys. Space Sci.* **321**: 69-72, (2009)